\documentclass[12pt,preprint]{aastex}
\usepackage{amsmath}

\begin{document}

\title{In-orbit demonstration of X-ray pulsar navigation with the \emph{Insight}-HXMT satellite}
\author{S. J. Zheng$^{1}$, S. N. Zhang$^{1,2}$, F. J. Lu$^{1,2}$, W. B. Wang$^{3}$, Y. Gao$^{3}$, T. P. Li$^{1,2,4}$, L. M. Song$^{1,2}$, M. Y. Ge$^{1}$, D. W. Han$^{1}$, Y. Chen$^{1}$
% ...
}

\affil{$^{1}$ Key Laboratory of Particle Astrophysics, Institute of High Energy Physics, Chinese Academy of Sciences, Beijing 100049, China; zhengsj@ihep.ac.cn, zhangsn@ihep.ac.cn}
\affil{$^{2}$ University of Chinese Academy of Sciences, Chinese Academy of Sciences, Beijing 100049, China}
\affil{$^{3}$ Technology and Engineering Center for Space Utilization Chinese Academy of Sciences, Beijing 10094, China}
\affil{$^{4}$ Department of Physics, Tsinghua University, Beijing 100084, China}
\affil{$^{5}$ Department of Astronomy, Beijing Normal University, Beijing 100088, China}
\affil{$^{6}$ Department of Engineering Physics, Tsinghua University, Beijing 100084, China}

\begin{abstract}

In this work, we report the in-orbit demonstration of X-ray pulsar navigation with \emph{Insight}-Hard X-ray Modulation Telescope (\emph{Insight}-HXMT), which was launched on Jun. 15$^{\rm th}$, 2017. The new pulsar navigation method `Significance Enhancement of Pulse-profile with Orbit-dynamics' (SEPO) is adopted to determine the orbit with observations of only one pulsar. In this test, the Crab pulsar is chosen and observed by \emph{Insight}-HXMT from Aug. 31$^{\rm th}$ to Sept. 5$^{\rm th}$ in 2017. Using the 5-day-long observation data, the orbit of \emph{Insight}-HXMT is determined successfully with the three telescopes onboard $-$ High Energy X-ray Telescope (HE), Medium Energy X-ray Telescope (ME) and Low Energy X-ray Telescope (LE) $-$ respectively. Combining all the data, the position and velocity of the \emph{Insight}-HXMT are pinpointed to within 10\,km (3$\sigma$) and
10 m\,s$^{-1}$ (3$\sigma$), respectively.

\end{abstract}


\section{Introduction}

Space navigation plays an important role for spacecraft launched to the earth orbit, lunar and Mars vicinities, and deep space. At present, the navigation of the spacecraft in low earth orbits mainly relies on global navigation satellite systems (GNSS) and the ground-based tracking systems, while the navigation in deep-space is mainly based on radio technologies (e.g. the US deep-space network). With the increasing number of space tasks, the ground stations are becoming overloaded and the operating costs are getting higher. Navigation also can not be handled timely, due to the communication delay or unexpected malfunction of ground stations. Pulsar navigation, an autonomous navigation technology, is receiving more and more attention as it is less dependent on the support of ground equipments and meets the continuous navigation requirements for space missions in different orbits.

The first pulsar was discovered in 1967 \citep{hxmtpn:Hewish}, and up until today more than 2000 pulsars have been discovered with frequency covering from radio, infrared, optical, ultraviolet, X-ray to gamma-ray. They are called `celestial GPS satellites' because of their long-term timing stability which is comparable to atomic clocks on the earth \citep{hxmtpn:Taylor}. The pulsar navigation using the radio and X-ray pulsars was proposed in 1974 \citep{hxmtpn:Downs} and in 1981 \citep{hxmtpn:Chester}, respectively. A large number of researches on the theory, algorithms and simulations have been carried out for spacecrafts in earth orbits, Mars orbits and deep space orbits (e.g. \cite{hxmtpn:Sheikh2, hxmtpn:Hanson, hxmtpn:EMADZADEH, hxmtpn:Becker, hxmtpn:Wei, hxmtpn:Wangyd, hxmtpn:Shemar, hxmtpn:Cui}), and the tracking phase or positioning errors for different pulsars are predicted in detail \citep{hxmtpn:Sheikh2, hxmtpn:Hanson}. In 1999, the first in-orbit test was carried out with the Unconventional Stellar Aspect (USA) on the ARGOS satellite \citep{hxmtpn:Wood, hxmtpn:Sheikh}. In 2016, the results of testing pulsar navigation with Gamma-Ray Bursts Polarimeter (POLAR), which was launched onboard the Chinese space laboratory TG-2 on Sept. 15$^{\rm th}$, 2016, were reported \citep{hxmtpn:Zhengsj}. With the effective area of $\sim$ 200\,cm$^2$ and wide field of view of more than 2$\pi$ Sr \citep{hxmtpn:Produit, hxmtpn:Lizhengheng}, POLAR has monitored the Crab pulsar for a long time. With 31-day-long observations of the Crab pulsar, the TG-2 orbit was determined successfully. The orbital elements were determined with the orbit deviation within 20\,km\,(1\,$\sigma$). On Jun. 3$^{\rm th}$, 2017, the Neutron Star Interior Composition Explorer (NICER) was launched on the International Space Station (ISS). It comprises 56 identical X-ray telescopes with effective area of 2000 \,cm$^2$ in total, each of the telescopes consisting of a concentrating X-ray optic and a single-pixel silicon drift detector\,(SDD) \citep{hxmtpn:Paul}. By measuring the Time of Arrival (TOAs) of five millisecond pulsars, NICER pinpointed its location within 5\,km \citep{hxmtpn:Alexandra}.

On Jun. 15$^{\rm th}$, 2017, \emph{Insight}-Hard X-ray Modulation Telescope (\emph{Insight}-HXMT) was launched in China \citep{hxmtpn:Zhangshu, hxmtpn:Lixiaobo}. It consists of three X-ray slat-collimated telescopes: High Energy X-ray Telescope (HE), Medium Energy X-ray Telescope (ME) and Low Energy X-ray Telescope (LE). HE contains 18 cylindrical NaI(Tl)/CsI(Na) Phoswich detectors with the energy band 20-250 keV and a total geometrical area of about 5000\,cm$^2$. ME is composed of 1728 Si-PIN detectors with the energy band 5-30\,keV and a total geometrical area of about 952\,cm$^2$. LE is composed of 96 Swept Charge Device (SCD) detectors which are sensitive in the 1-15\,keV with a total geometrical area of about 384\,cm$^2$. Due to the wide energy band and large detection area, \emph{Insight}-HXMT can also be used to make in-orbit test of pulsar navigation. In this paper, the new pulsar navigation method `Significance Enhancement of Pulse-profile with Orbit-dynamics' (SEPO), which has been tested in POLAR \citep{hxmtpn:Zhengsj}, is adopted to determine the orbit of \emph{Insight}-HXMT autonomously. The Kalman Filter method can also be used to make in-orbit demonstration of X-ray pulsar navigation \citep{hxmtpn:Wangyd2}.

\section{Navigation method}
Conventionally, at least three pulsars are needed in pulsar navigation models to estimate the position of a spacecraft. However, it results in complexity of the payload and poses risks for control system. The  one-pulsar navigation method SEPO has been proposed by combining the significance analysis of pulse profile and orbit dynamics. The orbital dynamics model can produce a high-precision orbit forecast in a short period of time but with a long-term drift. The pulse profile will be `deformed' due to the orbit deviation as explained below, resulting in the decrease of the significance of the profile signal, which is used to update the orbital parameters continuously.

\subsection{The orbit dynamics}\label{orbit_dynamics}

The orbit is described by six orbital elements: semi-major axis ($a$), eccentricity ($e$), inclination angle ($i$), right ascension of the ascending node (RAAN, $\omega$), argument of perigee ($\Omega$), and mean anomaly ($w$). Considering the altitude of the \emph{Insight}-HXMT orbit (about 550\,km) and the complexity of the earth's perturbation force, the orbit forecast model is constructed with the 70$^{\rm th}$ order gravity field, the atmospheric drag, and the solar radiation pressure, and the track integrator is Runge-Kutta-Fehlberg 7(8) with an integration step of 60 seconds \citep{hxmtpn:Wangwb}. In such conditions, the position error is less than 3.5\,km for the forecast orbit in five days. It could be further simplified according to the different conditions of a spacecraft.

\subsection{The significance of the pulse profiles}

To get the `standard' pulse profile, the arrival time of each photon on the local site is corrected to the Solar System Barycenter (SSB) and folded with the pulsar ephemeris with certain bins \citep{hxmtpn:Gemy}. In this process, the orbit of the spacecraft plays an important role for the time correction. If the orbit deviates from the correct position, the calculated profile will be deformed from the `standard' profile due to the wrong phase.

The significance of the pulse profile is defined as follows;

\begin{equation}
\chi^2=\sum_{i=1}^{n}\frac{({P(\phi_i)-\bar{P}})^2}{\bar{P}},
\end{equation}
where $P(\phi_i)$ is the counts of the profile at $\phi_i$ , $\bar{P}$ is the mean counts of the profile and $n$ is the total bin number of the profile.

It is expected that the significance of the calculated profile varies with spacecraft orbit deviation. The more the orbit deviates from the real one, the more the profile will be deformed, and the less the significance. $\chi^2$ will reach the maximum for the zero deviation, i.e., the true orbit.

\begin{figure}[!h]
\centering
\includegraphics[width=0.8\textwidth]{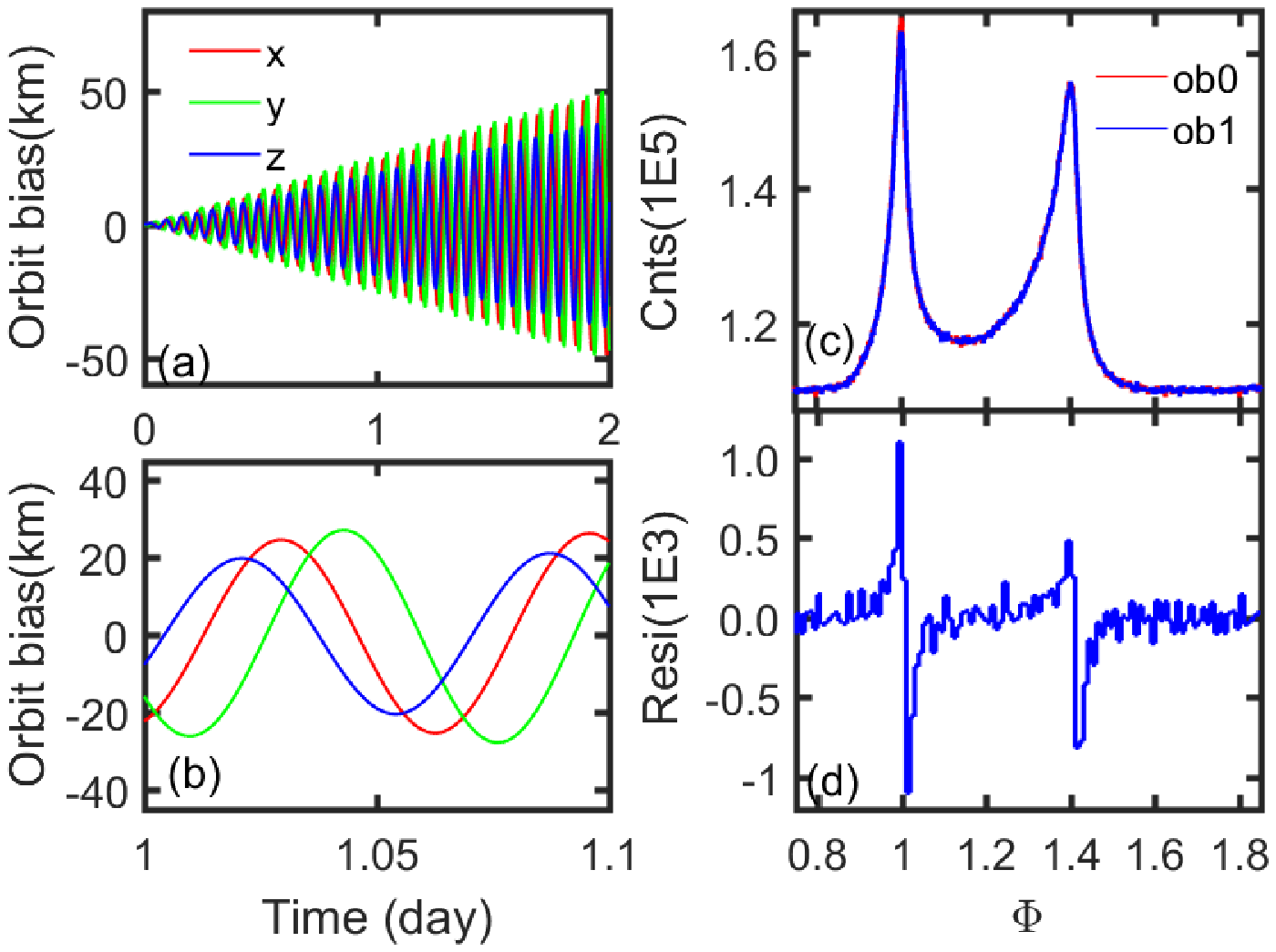}
\caption{The orbit deviation due to the change of initial orbit element `semi-major axis' by 200 m and its effects on the profile. (a) orbit deviations for 2 days; (b) zoom in of the orbit deviations. (c) distortion of pulse profile induced by the change of initial element. The red line represents the `standard' profile generated with `ob0', and the blue one represents the distorted profile generated with `ob1'; (d) differences between the `standard' profile and distorted profile}
\label{fig_sim}
\end{figure}

We make a simulation for five-day observations of the Crab pulsar with HE. The true orbit `ob0' is given by the GPS on the satellite, and the orbit `ob1' is calculated with the dynamics of orbits by changing the `semi-major axis' with 200\,m. As shown in Fig. \ref{fig_sim}, the deviation will be larger, to more than 50\,km after 2 days. Using the predicted orbit `ob1', the profile is deformed and the significance is decreased from $1.237\times10^6$ to $1.226\times10^6$, i.e. by about $1\%$.

%
%

\subsection{The orbit fitting with grid search}

\begin{figure}[!h]
\centering
\includegraphics[width=0.4\textwidth]{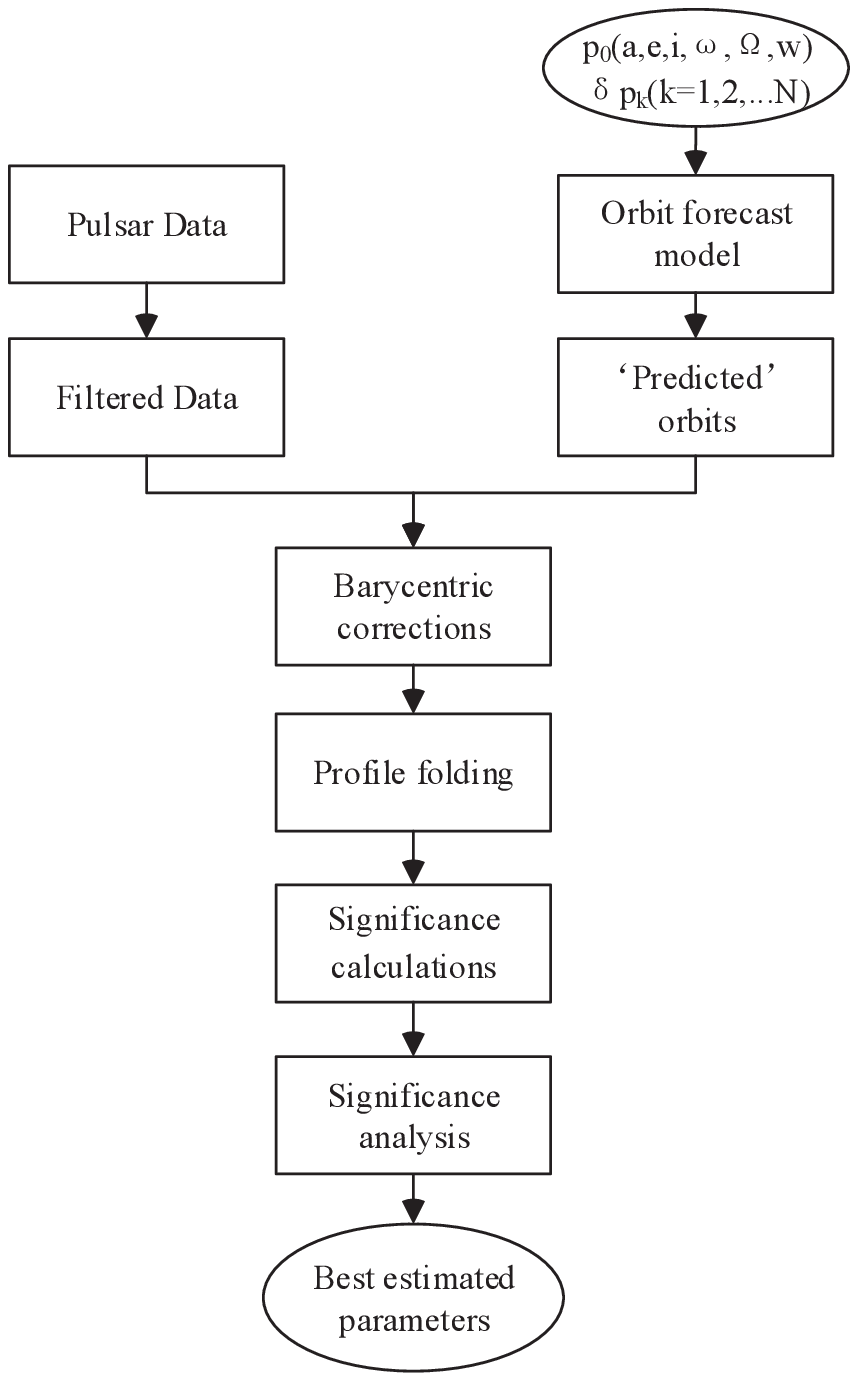}
\caption{Processing flow of pulsar navigation.}
\label{algorithm}
\end{figure}

Considering the correlations between the significance of the profile and the orbit deviation, the orbit determination algorithm is described by the processing flow in Fig. \ref{algorithm}. In this algorithm, the orbit forecast and the profile's significance analysis are combined. The predicted orbit is obtained with orbital dynamics model to complete the profile folding. At the same time, the long-term stability characteristics of the profile can be used to continuously update the orbital parameters, thus the long-term drift of the orbit model can be corrected.

On one hand, the orbit forecast model is constructed by considering the orbit dynamics as described in Section \ref{orbit_dynamics}. Then, with the estimated orbit parameters $p_0$\,($a$, $e$, $i$, $\omega$, $\Omega$, $w$) at the current time ($t_0$) and the certain parameter spaces $\delta p_k\,(k=1,2,..N)$, the $k^{\rm th}$ predicted orbit is generated for the time span [$t_s$, $t_0$] ($t_s$ is the beginning time for the used historic observation data). On the other hand, the used data with the time span [$t_s$, $t_0$] are chosen and filtered according to the selection criteria as described in Section \ref{data_reduction}.

With the predicted orbit at different grid point, the arrival times of photons are corrected to SSB and folded to generate the `predicted' profiles using phases calculated with the following equation \citep{hxmtpn:Gemy};

\begin{equation}
\Phi-\Phi_{0}=\nu{(t-t_{e})}+\frac{1}{2}\dot\nu{(t-t_{e})}^{2}+\frac{1}{6}\ddot\nu{(t-t_{e})}^{2},
\label{phase_eq}
\end{equation}
where $\Phi_{0}$, $\nu$, $\dot\nu$ and $\ddot\nu$ are the phase,  frequency, frequency derivative and the second order derivative of frequency at the reference epoch $t_{e}$, respectively. Then, all the significances for different profiles are calculated. Finally, the `significance analysis' is carried out, i.e. the trend between the significance of profiles and the orbit parameters is fitted with the Gaussian function and the optimal orbit parameters are obtained.

\section{Data reduction}\label{data_reduction}

Almost five-day observation data of the Crab pulsar from 2017-08-31 10:00:00 to 2017-09-05 08:00:00 (UTC) are chosen, including all the data collected by HE, ME, and LE. The data are analyzed with the \emph{Insight}-HXMT Data Analysis software (HXMTDAS) software v2.0\footnote{http://www.uu-world.cn/hxmt/hxmtsoft/index2.html}. The following data selection criteria as recommended in the software are used:

(1) Pointing offset angles $\leq 0.05^\circ$;

(2) Elevation angles $\geq 12^\circ$;

(3) Geomagnetic cutoff rigidity $\geq 6$;

(4) Only the detectors with small FoVs selected.

In addition to the above criteria, intervals with high background induced by the charged particles or bright earth are excluded manually. Then we get the filtered data with the total exposure duration of 180\,ks, 190\,ks, and 120\,ks for HE, ME, LE, respectively. The arrival times of the photons are corrected to SSB, then the pulsed profiles are folded for HE, ME, and LE. The pulsar ephemeris, shown in Table \ref{psrpar}, is obtained from \emph{Insight}-HXMT data with the timing residual of about 10 $\mu$s (1$\sigma$). As the $\ddot{\nu}$ parameter of the Crab pulsar is about $10^{-20}\,{\rm s}^{-3}$, the affected pulsar phase offset after 5 days is negligible ($\frac{1}{6}\cdot\ddot{\nu}\cdot dt^3 =1.3\times10^{-4}$), thus we set $\ddot{\nu}=0$.


\begin{table}[htb]
\renewcommand{\arraystretch}{1.3}
\caption{The pulsar parameters of the Crab pulsar}
\label{psrpar}
\centering
\begin{tabular}{cc}
\hline
RA & $05^{\rm h}34^{\rm m}31^{\rm s}.973$\\
DEC & $22^\circ00'52.06''$\\
$\nu$ & 29.639022542326\\
$\dot\nu$ & $-3.6867\times10^{-10}$\\
$\ddot\nu$ & 0\\
EPOCH & 57992.16548081821\\
%
\hline
\end{tabular}
\end{table}

\section{Results}


The orbital elements of \emph{Insight}-HXMT are given at 07:59:00 (UTC) on Sept. 5$^{\rm th}$, 2017, and the space grids of parameters are chosen as listed in Table \ref{orbpar}. For an easy display, only the deviations of the parameters ($a$, $e$, $i$, $\omega$, $\Omega$, $w$) from the true values ($6922.8781$\,km, $0.00181017$, $42.9715^{\circ}$, $207.0229^{\circ}$, $116.9049^{\circ}$, $22.5215^{\circ}$) are listed.

\begin{table}[htb]
\renewcommand{\arraystretch}{1.3}
\caption{The orbit parameter spaces. '$\Delta a$', '$\Delta e$', '$\Delta i$', '$\Delta\omega$', '$\Delta\Omega$' and '$\Delta w$' are the deviations for semi-major axis, eccentricity, inclination angle, RAAN, argument of perigee, and mean anomaly, respectively.}
\label{orbpar}
\centering
\begin{tabular}{cccc}
\hline
Orbital element deviation & Min & Max & Step \\
\hline
$\Delta$\,a\,(m) & -200 & 200 & 20 \\
$\Delta$\,e\qquad & -0.0018 & 0.0182 & 0.001 \\
$\Delta$\,i\,($^\circ$) & -0.5 & 0.5 & 0.05 \\
$\Delta$\,$\omega$\,($^\circ$) & -1.0 & 1.0 & 0.1 \\
$\Delta$\,$\Omega$\,($^\circ$) & -1.0 & 1.0 & 0.1 \\
$\Delta$\,w\,($^\circ$) & -1.0 & 1.0 & 0.1 \\
\hline
\end{tabular}
\end{table}

\begin{figure}[htb]
\centering
\includegraphics[width=0.7\textwidth]{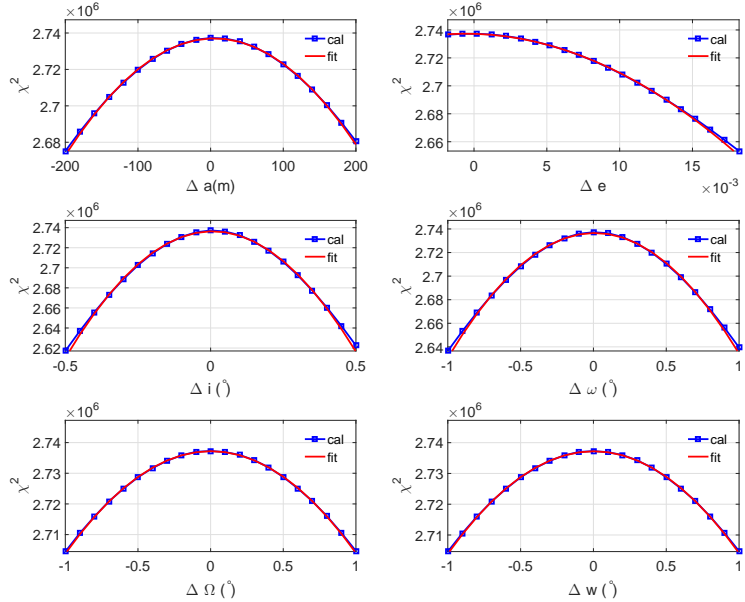}
\caption{Orbit determine result with Insight/LE: $\chi^2$ of the orbit elements of calculations (blue squares) and fitted results (red lines). '$\Delta$a', '$\Delta$e', '$\Delta$i', '$\Delta$$\omega$', '$\Delta$$\Omega$' and '$\Delta{w}$' are the deviations for semi-major axis, eccentricity, inclination angle, RAAN, argument of perigee, and mean anomaly, respectively.}
\label{LEnav}
\end{figure}

First, the orbit determination method is demonstrated with LE. At each grid point of the parameter space, the `predicted' profile is folded and its significance is obtained.
As shown in Fig. \ref{LEnav}, the significance of the profile varies significantly with the deviation of the orbital parameters and has a maximum around the true value (zero deviation). Then the maxima obtained by fitting a Gaussian function curve give the optimal values for the six orbital parameters separately, i.e. each time one parameter is free with the other five fixed, and the results are shown in Table \ref{deviation}. The process obtaining the best parameters is very similar to that used by Sheikh et al to get the TOAs from the observed pulse profiles \citep{hxmtpn:Sheikh}. Therefore, the uncertainties of these parameter values can be obtained in a similar way as follows;

\begin{equation}
\sigma_p=\frac{\frac{1}{2}W}{SNR},
\end{equation}
where {\em W} is Full Width Half Maximum (FWHM), and {\em SNR} is the signal to noise ratio obtained by the bootstrap method \citep{hxmtpn:Diaconis}. As mentioned by Sheikh \citep{hxmtpn:Sheikh}, the SNR is limited to a maximum of 1000 by:

\begin{equation}
SNR_{\rm filtered}=\frac{1000SNR}{1000+SNR}.
\end{equation}


Then similar demonstrations are also done for HE and ME, shown in Table \ref{deviation}. It shows that the errors for HE are less than for ME and LE, because it has collected more pulsed photons. Combining all the data from HE, ME and LE, the errors become smaller. Transfer the orbit elements to the cartesian parameters, we obtain the 3$\sigma$ errors of the current position
and velocity, which are 3.15\,km, 6.61\,km, and 4.46\,km for {\em x, y, z} and 0.0073\,km\,s$^{-1}$, 0.0033\,km\,s$^{-1}$, and 0.0042\,km\,s$^{-1}$ for $v_x$, $v_y$, $v_z$, respectively. In short,  the position and velocity are pinpointed within 10\,km (3$\sigma$) and 10 m\,s$^{-1}$ (3$\sigma$).

\begin{table}[h]
\renewcommand{\arraystretch}{1.3}
\caption{Best estimated value of the elements and  errors (3$\sigma$)}
\label{deviation}
\centering
\begin{tabular}{cccccccc}
\hline
Payload & Orbital element deviation & $\Delta$a & $\Delta$e & $\Delta$i & $\Delta\omega$ & $\Delta\Omega$ & $\Delta$w \\
   &   & m & $10^{-3}$ & $^\circ$ & $^\circ$& $^\circ$& $^\circ$ \\
\hline
LE & deviation &7.02&-0.38&0.01&0.01&0.003&0.003\\
   & error ($3\sigma$)&11.75&1.03&0.02&0.06&0.09&0.09 \\
ME & deviation &3.07&-0.11&0.005&0.01&0.007&0.006\\
   & error ($3\sigma$)&16.44&1.28&0.03&0.064&0.11&0.11 \\
HE & deviation &4.89&-0.58&0.006&0.009&0.002&0.002\\
   & error ($3\sigma$)&7.58&0.59&0.01&0.03&0.05&0.05 \\
all detectors & deviation &5.25&-0.50&0.007&0.01&0.002&0.002\\
   & error ($3\sigma$)&6.40&0.52&0.01&0.03&0.04&0.04 \\
\hline
\end{tabular}
\end{table}


\section{Discussions}
Conventionally, a minimum of three pulsars are required to get the absolute navigation. Considering the clock time-offset on the spacecraft, another pulsar is needed. Thus at least four pulsars should be observed simultaneously, which would require at least four detectors, increasing the technical complexity and the cost of mass, energy and money. An alternate choice is to observe different pulsars sequentially using a single detector, which would increase the risk of the control system. NICER has adopted the latter strategy. With the large effective areas and low background rates, it could get the TOAs of some millisecond pulsars in several kiloseconds. By observing five millisecond pulsars sequentially, the spacecraft position was determined under real flight conditions successfully \citep{hxmtpn:Alexandra}.

The method SEPO, determining the orbit by using only one pulsar, has been proposed and demonstrated with POLAR \citep{hxmtpn:Zhengsj} and \emph{Insight}-HXMT in this paper, respectively. As POLAR has an effective area of about 200\,cm$^2$ and high background rates (about 4000 counts\,s$^{-1}$), the orbit was determined within 20\,km (1\,$\sigma$) by monitoring the Crab pulsar for 1 month. However, it should be pointed out that the POLAR's response for photons varies with their incident angles, thus the obtained different profiles are mixed and become `broader', decreasing the navigation precision which is shown in appendix A. For \emph{Insight}-HXMT, with the pointed observation mode, the response to the incident pulsar emission is almost the same. With the observations of the Crab pulsar for 5 days, the position is determined within 10\,km. So it is feasible to use one detector for navigation by monitoring one pulsar for a long time, which would result in low request of mass, energy, and less spacecraft control. It is favorable for the navigation applications, particular for the deep space navigation during the cruise phase of flight.

In addition, the standard pulse profile and parameters are important external inputs in X-ray pulsar navigation. However, it is hard to achieve the fixed `standard' profile in space X-ray detection. For example, if the energy band or the response of the instrument is different, the `standard' profiles become different. Furthermore, the response of the instrument may vary in the long-term operation in space, thus the profile varies too. If we still use the previous `standard' profile, it would inevitably affect the navigation accuracy and even result in unpredictable systematic deviation. For SEPO, neither the standard pulse profile nor the continuous update of the pulsar parameters is needed. Therefore, it is no longer dependent on the standard pulse profile, nor affected by the pulsar observation conditions and the detector performance changes. It can be applied to different detectors and different mission scenarios.

For the in-orbit detections of a pulsar, the obtained profile could be regarded as the summation of all the profiles at different point of the orbit. That is, $P(\phi_i)=\sum_{k=1}^{m}p_k(\phi_i)=\sum_{k=1}^{m}p(\phi_i-\delta{\phi_k})$,
where $\delta{\phi_k}$ represents the phase offset due to the position deviation. Thus, considering the significance of the profile defined in Sec. 2.2,

\begin{equation}
\begin{aligned}
\chi^2 &=\sum_{i=1}^{n}\{{\sum_{k=1}^{m}p(\phi_i-\delta{\phi_k})-\bar{P}}\}^2/\bar{P}\\
&=\sum_{i=1}^{n}\sum_{k=1}^{m}\sum_{k'=1}^{m}p(\phi_i-\delta{\phi_k})p(\phi_i-\delta{\phi_{k'}})/\bar{P}-n\bar{P}.
\end{aligned}
\label{prochi}
\end{equation}

As shown in equation \ref{prochi}, $\chi^2$ of the profile will reach the maximum with $\delta\phi_k=\delta\phi_{k'}$, i.e. $\phi_k=const$.  $\phi_k=0$ represents the true orbit with zero deviation. If $\phi_k=const$ (nonzero), it means that the orbit has fixed deviations from the real orbit. However, constrained by the orbital dynamics, there is few such completely `parallel' tracks, thus ensuring to find the true orbit. For the cruising orbit in deep space, there indeed exist approximately parallel tracks. In such conditions, more measurements besides the significance of the profiles, TOAs for example, are needed to discriminate the true orbit from the others.

The proposed SEPO method, however, has not been proven mathematically. Thus some simulations have been performed and the detailed results are shown in the appendix, which shows that the method works with very different pulse profiles. In this in-orbit demonstration with the SEPO method, the position is pinpointed to within 10\,km (3$\sigma$), while the theoretical bound is $\sim$ 250 meters (3$\sigma$) for 1920\,m$^2\cdot$s of LE (the effective area of 160\,cm$^2$ for the small FOV with the exposure duration of 120\,ks) \citep{hxmtpn:Hanson}. Possible reason are the effects of orbit forecast error ($<3.5$\,km), absolute time accuracy of \emph{Insight}-HXMT \citep{hxmtpn:Lixiaobo}, or the intrinsic timing noise of the Crab pulsar. The ephemeris used in this demonstration has an timing residual of $\sim$10\,$\mu$s (1$\sigma$), resulting in position error of $\sim$9.0\,km (3$\sigma$). See the detailed discussions on the effects of timing noise at the end of this section.


By now, the clock time-offset on the spacecraft has not been considered in the SEPO method. The short-term drift of clock within the integration time would deform the calculated profile, change the pulse significance and result in additional systematic errors. While the long-term drift just changes the absolute TOA, which has little effects on the pulse significance. Thus the clock time-offset can be taken into account and corrected with the significance analysis in future.
In addition, the pointing error also decreases the significance of the profiles by changing the effective areas. However, the pointing error has no effects on the pulse phase of the pulsar, thus has little effect on the solution in the SEPO method, even if the pointing error has a dependence on orbit position. In this work, the observation of \emph{Insight}-HXMT was performed in the `pointed mode' with the pointing error $\leq 0.028^\circ$ (3$\sigma$). In this situation, the effective area changes by only about 0.5\%, which can be ignored completely.

The timing noise of pulsars \citep{hxmtpn:Hobbs} has a great influence on the accuracy of pulsar navigation as shown above. The timing noise of a pulsar includes white noise component and red noise component. White noise can be suppressed by accumulating observation data over a long period of time. However, the red noise is difficult to suppress or eliminate. Actually, the effects of the red noise is quite similar to the clock time-offset and can be corrected as mentioned above. On the other hand, combining with the study of the mechanism of noise (such as magnetic field evolution), it is possible to give a more accurate prediction of the spin evolutions of pulsars, reducing the impact of timing noise (especially red noise) on the accuracy of pulsar navigation \citep{hxmtpn:Yisx, hxmtpn:Gaoxd, hxmtpn:Zhangsn1, hxmtpn:Zhangsn2}. In addition, the glitch, that is, a sudden change in the rotation period, will also have to be considered in future.

\section{Conclusion}
A new navigation method SEPO has been proposed that combines the observed pulse profile with orbit dynamics. It has been demonstrated with the \emph{Insight}-HXMT  observations of only one pulsar (Crab) and the orbit has been determined successfully. Combining all the data, we obtain the best position estimations within 10\,km (3$\sigma$) and velocity within 10 m\,s$^{-1}$ (3$\sigma$), respectively.

\acknowledgments{This work is supported by the National Key R\&D Program of China (2016YFA0400800) and the National Natural
Science Foundation of China under grants U1838101 and 11503027. This work made use of the data from the \emph{Insight}-HXMT mission, a project funded by China National Space Administration (CNSA) and the Chinese Academy of Sciences (CAS).}

\clearpage

\appendix{Appendix}

\section{Simulations on the SEPO method}\label{appendix}
We perform some simulations to evaluate the feasibility of the new navigation method SEPO, especially for different pulse-profiles. The duty cycle of the profile $\delta$ is defined as follows;

\begin{equation}
\delta=\frac{w}{d},
\end{equation}
where $w$ is the width of the pulse (above 10\% maximum of the peak), $d$ is the pulsar period. Thus $\delta$ describes the sharpness of a pulse profile.

\begin{figure}[htb]
\centering
\includegraphics[width=0.7\textwidth]{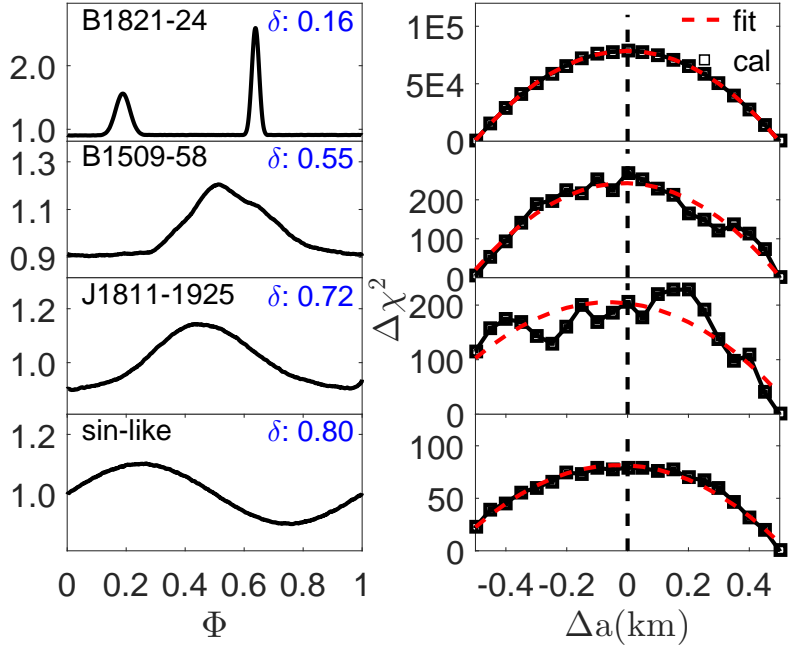}
\caption{Left: the four pulse-profiles used in the simulations, `$\delta$' represents the duty cycle as defined in the text; right: $\Delta\chi^2$ of the deviations for semi-major axis ($\Delta$a) of calculations (black squares) and fitted results (red lines). For clarity, $\Delta\chi^2$ ($\chi^2-\chi^2_{\rm min}$) is shown. $\chi^2_{\rm min}$ is different for different pulse-profiles, which is $7.09\times10^6$, $1.00\times10^6$, $7.22\times10^5$, $5.05\times10^5$ for PSRs B1821-24, B1509-58, J1811-1925 and sin-like profile, respectively. }
\label{LEsim}
\end{figure}

First, three different pulse profiles from three pulsars, i.e., PSRs B1821-24, B1509-58, J1811-1925 and a supposed sinusoidal-like pulse shape are chosen for test, as shown in Fig. \ref{LEsim} (the left panel). To exclude the effects of other factors, the parameters such as position and period are fixed to be the same as the Crab pulsar. The pulse count rate is assumed to be 130 cnts$\cdot$ s$^{-1}$ and the background count rate to be 1200 cnts$\cdot$ s$^{-1}$, as obtained from the LE data. Then the \emph{Insight}-HXMT/LE observation data of the four pulsars are simulated with the true orbit of \emph{Insight}-HXMT from Aug. 30$^{\rm th}$ to Sept. 5$^{\rm th}$, 2019.

Then the method SEPO is carried out and the significance of the profiles with the deviations of the elements (semi-major axis for example) are shown in Fig. \ref{LEsim} (the right panel). It shows that the significance of all the four different pulse-profiles varies with $\Delta$a, and the optimal value can be found by Gaussian-fitting, which is near the true value (zero). The pulse-profiles do have effects on the position error. The sharper is the pulse-profile, the larger is the calculated significance, and the smaller is the position error. The duty cycle $\delta$ is 0.16, 0.55, 0.72, 0.80 for PSRs B1821-24, B1509-58, J1811-1925 and sinusoidal-like pulsar. The profiles like PSR B1821-24 are the best candidate due to its the navigation for the small duty cycle. For PSR J1811-1925, the significance trend is not smooth, however, the optimal value can still be obtained by fitting.

\begin{figure}[htb]
\centering
\includegraphics[width=0.7\textwidth]{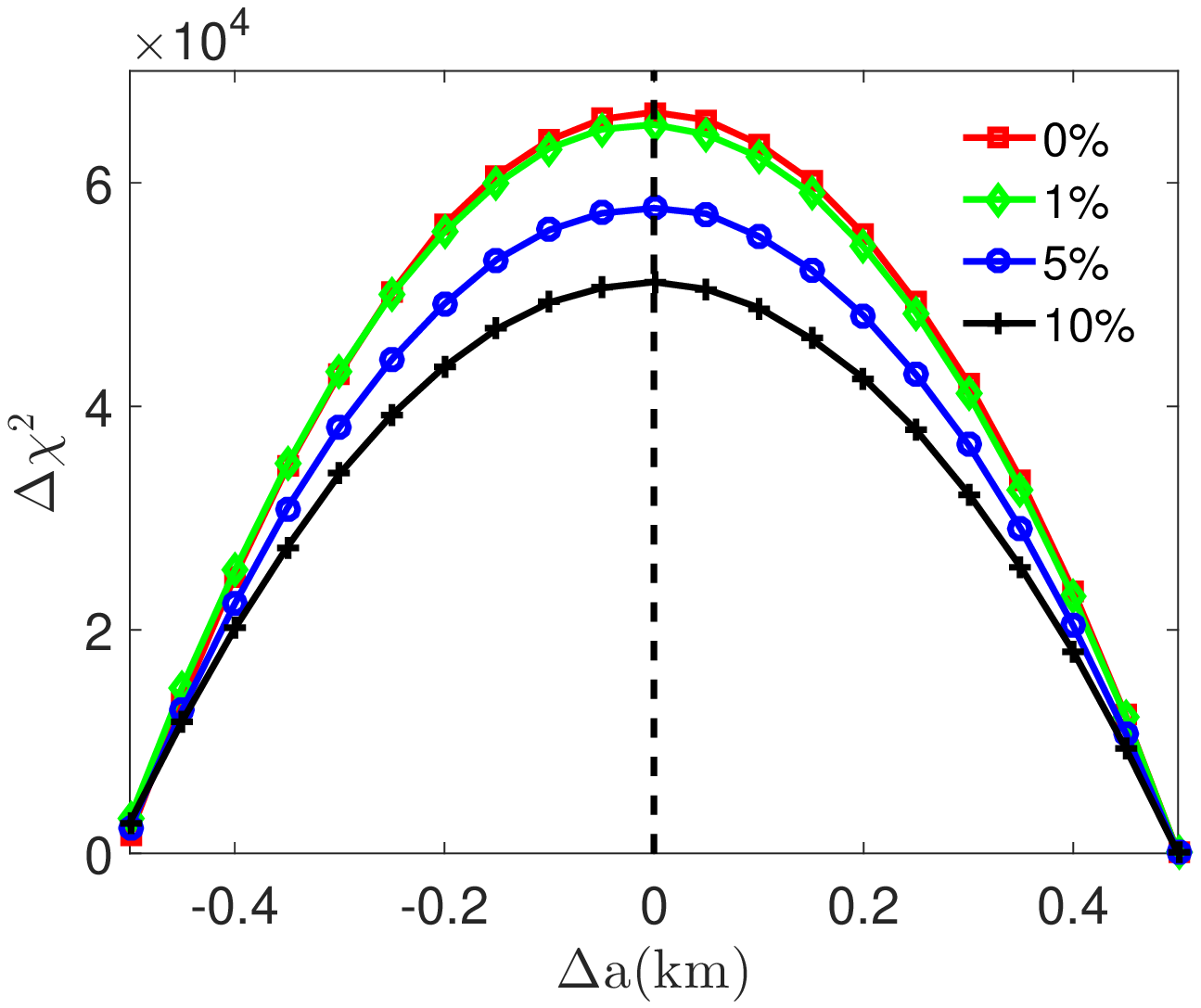}
\caption{$\Delta\chi^2$ of the deviations for semi-major axis ($\Delta$a) of calculations (black squares) and fitted results (red lines). For clarity, $\Delta\chi^2$ ($\chi^2-\chi^2_{\rm
min}$) is shown. $\chi^2_{\rm min}$ is different for different pulse-profiles, which is $6.40\times10^6$, $6.33\times10^6$, $6.06\times10^6$, $5.75\times10^6$ for the changed profile with $\delta$ increased by 0\%, 1\%, 5\%, and 10\%, respectively. }
\label{pro_cmp}
\end{figure}

To estimate the effects of pulse shapes in details, a double-peak profile is generated with $\delta=0.16$, and $\delta$ is increased by 1\%, 5\%, and 10\%, respectively. Then the observation data are simulated and the method SEPO is carried out as described above. The results are shown in Fig. \ref{pro_cmp}, and the optimal semi-major axis and the errors ($3\sigma$) are shown in Table \ref{sim_deviation}. It also shows that the significance decreases as the profile gets broader. With $\delta$ becoming larger by 10\%, the resulted error of the semi-major axis is increased by 8.8\%.

\begin{table}[h]
\renewcommand{\arraystretch}{1.3}
\caption{Best estimated value of $\Delta$a and  errors (3$\sigma$)}
\label{sim_deviation}
\centering
\begin{tabular}{cccccc}
\hline
enlargement of $\delta$ & & 0\% & 1\% & 5\% & 10\% \\
\hline
$\Delta$a (m) & deviation & -3.15 & -6.12 & -5.4 & -6.8 \\
              & error ($3\sigma$) & 24.96 & 25.12 & 26.12 & 27.15 \\
\hline
\end{tabular}
\end{table}

\end{document}